\documentclass{pasj01}

\let\leftrightarrows\relax
\let\gtrsim\relax

\let\lesssim\relax

\let\leftroot\relax
\let\uproot\relax

\let\varnothing\relax

\Received{$\langle$ $\rangle$}
\Accepted{$\langle$ $\rangle$}
\Published{$\langle$ $\rangle$}
\SetRunningHead{Astronomical Society of Japan}{Usage of \texttt{pasj00.cls}}
\usepackage{amsmath, amssymb}
\usepackage[utf8]{inputenc}   
\usepackage[T1]{fontenc}      
\usepackage{textcomp}         

\newcommand{\dodoi}{DODOI}
\newcommand{\rfwhm}{R_{\rm{FWHM}}}
\newcommand{\rbar}{\bar{R}} 
\newcommand{\rsigma}{\sigma_{R}}
\newcommand{\wb}{7.23\times 10^{-1}~\rm{\mu G}}
\newcommand{\mb}{2.16~\rm{\mu G}}
\newcommand{\stb}{7.23~\rm{\mu G}}

\usepackage{booktabs}
\usepackage{multirow}
\usepackage{threeparttable}
\usepackage{graphicx}

\begin{document}

\title{A statistical approach for interpreting polarized dust emission of the filamentary molecular clouds toward the estimate of 3D magnetic field structure}
\author{Haruka Fukihara,\altaffilmark{1} \footnotemark[*]
        Daisuke Takaishi,\altaffilmark{1}
        Yoshiaki Misugi,\altaffilmark{2}
        Megumi Sasaki,\altaffilmark{1} and
        Yusuke Tsukamoto\altaffilmark{1}}
\altaffiltext{1}{Graduate School of Science and Engineering, Kagoshima University, 1-21-35 Korimoto, Kagoshima, Kagoshima 890-0065, Japan}
\altaffiltext{2}{Division of Science, National Astronomical Observatory of Japan, 2-21-1 Osawa, Mitaka, Tokyo 181-8588, Japan}
\email{k9786351@kadai.jp}

\maketitle

\begin{abstract}
  In this study, we perform 3D magnetohydrodynamics (MHD) simulations of filamentary molecular clouds. We then generate synthetic observations based on the simulation results. Using these, we investigate how the new polarization data analysis method recently introduced by \citet{2021ApJ...923L...9D} reflects the magnetic field structure in turbulent filamentary molecular clouds. 

  \citet{2021ApJ...923L...9D} proposed that the $\rfwhm$, the ratio of the Full Width at Half Maximum (FWHM) of the polarized intensity ($PI$) to that of the total intensity ($I$) can be used to probe the three-dimensional structure of the magnetic field.
  We calculate the $\rfwhm$ from the density and magnetic field structure obtained in the 3D-MHD simulations. We find that the mean and variance of $\rfwhm$ within a filament are smaller and larger, respectively, with a larger inclination of the magnetic field to the plane-of-sky.
  We also find that both small-scale ($<0.1~\rm{pc}$) and large-scale ($\gtrsim 0.1~\rm{pc}$) turbulence affect the polarized intensity of the dust thermal emission.
  We conclude that future extensive observations of $\rfwhm$ may be able to quantify the inclination of the magnetic field to the plane-of-sky in the filamentary molecular clouds.
\end{abstract}

\section{Introduction}
\label{sec_intro}
Observations by the Herschel Space Telescope show the universality of filamentary structures in molecular clouds\citep{2010A&A...518L.102A}.
The observed filaments have the common width of $\sim 0.1~\rm pc$ and subsequent studies have also shown the complex dynamics in the filament as well as that protostars are distributed along these filaments
\citep{2010A&A...518L.102A, 2010A&A...518L.100M, 2011A&A...529L...6A, 2013A&A...553A.119A,2013A&A...550A..38P, 2014ApJ...791...27S,2019A&A...621A..42A,2023ASPC..534..233P}.
These discoveries have stimulated considerable interest in unraveling the dynamics of the filaments and their role in the star formation processes
\citep{2016MNRAS.457..375F,2019ApJ...881...11M, 2024ApJ...963..106M}.

In order to understand the dynamics of molecular clouds and star formation within them, it is essential to know the role of magnetic fields, and the intensive observations of magnetic fields in filamentary molecular clouds have been made in the last decade.
In particular, BISTRO (B-Fields in Star-forming Region Observation) project \citep{2017ApJ...842...66W},
a survey project of the polarized dust thermal emission from molecular clouds, has revealed complex and diverse magnetic field structures inside molecular clouds
\citep{2018ApJ...859....4K,2019ApJ...877...43L,2021A&A...647A..78A,2021ApJ...912L..27E}.
\citet{2018ApJ...860L...6P} revealed the magnetic field structure inside the Pillars of Creation in M16 and pointed out the importance of the degree of coupling between the pillars and the surrounding ionized gas via the magnetic field for the evolution and lifetime of the pillars.

As observational sensitivity improves and observations provide more detailed information on the polarization direction and polarized intensity of dust thermal emission,
deciphering the observed data to understand the dynamical role of the magnetic field in molecular cloud evolution becomes increasingly challenging.
To address this challenge, \citet{2021ApJ...923L...9D} introduced a novel approach to interpret polarization data to obtain insights into the 3D magnetic field structure inside filaments. 
They introduced a scalar quantity, $\rfwhm$, defined as the ratio of the Full Width at Half Maximum (FWHM) of the polarized intensity to that of the total intensity along the minor axis of filaments, as an indicator to characterize the global magnetic field structure.

By utilizing the magnetohydrostatic filament model, they concluded that the observed small $\rfwhm$ in the NGC1333 filament is attributed to a significantly
pinched magnetic field toward its central axis. 
The findings of \citet{2021ApJ...923L...9D} are intriguing on their own, but their study also highlights the potential of utilizing simple scalar quantities to understand complex vector fields, i.e., 3D magnetic field structures in molecular clouds.

While their methodologies are great advancements, they rely on the magnetohydrostatic filament models for the interpretation of the observed data,
which may not fully capture the dynamical evolution of magnetic fields in realistic turbulent filaments.
Therefore, it is essential to further explore the significance of $\rfwhm$ in the dynamically evolving turbulent magnetized filaments.

The dynamical evolution of filaments has been a subject of theoretical study for decades.
Pioneering studies by \citet{1992ApJ...388..392I} and \citet{1997ApJ...480..681I} investigated the collapse and fragmentation processes of filaments through linear analysis and simulations. 
They revealed that quasi-equilibrium filaments fragment into dense cores with separations of approximately four times the filament diameter.
Subsequent theoretical studies using 3D models, such as numerical simulation by \citet{2015MNRAS.449L.123M} and analytical calculation by \citet{2019ApJ...881...11M}, have investigated the dynamical evolution of filaments and the properties of the cores formed within them.

\citet{2015MNRAS.449L.123M} conducted 3D simulations covering a wide range of dynamical scales, from filaments to protostars, and examined the detailed core evolution resulting from filament fragmentation.
The velocity structure obtained in their simulations is consistent with an ALMA observation in the inner region of a molecular cloud core \citep{2016ApJ...826...26T}.
\citet{2019ApJ...881...11M} focused on the role of turbulence in initiating filament fragmentation and explored its implications for the angular momentum evolution of cloud cores.
Their study revealed the crucial role of turbulence in shaping the properties of cloud cores born within filaments.

Despite theoretical advancements, the previous studies have ignored the magnetic fields on filament dynamics\footnote{
During the preparation of this paper, \citet{2024ApJ...963..106M} published a paper on the dynamics of magnetized filaments.

A significant difference between their work and ours lies in the focus: we focus on the observational characteristics of magnetic fields in the dynamically evolving filaments, while \citet{2024ApJ...963..106M} concentrate on the detailed dynamics of gas and magnetic fields.
}.
With increasing observational evidence emphasizing the importance of magnetic fields in molecular cloud evolution,
it would be required to integrate magnetic fields into the simulations of filament dynamics and elucidate their observational signatures.

In this study, by combining 3D magnetohydrodynamics (MHD) simulations and synthetic observations of dust polarized emission, we investigate the evolution of the magnetized turbulent filaments and explore how magnetic field structures are reflected in the features of observed polarized dust emission.

This paper is organized as follows. We describe the methods employed in this study in Section \ref{sec_method}.
Section \ref{sec_results} presents our findings, the results of the 3D-MHD simulation (Section \ref{subsec_sim_results}), the results of the synthetic observation (Section \ref{subsec_synob_results}), and 
the relation between the magnetic field structures obtained from the simulation and the polarization profiles obtained from the synthetic observation (Section \ref{subsec_relation_bw_BandP}).
Finally, we discuss the implications of our findings in Section \ref{sec_conclusion}.

\section{Method}
\label{sec_method}

\subsection{Numerical simulation}
For 3D magnetohydrodynamics (MHD) simulations of filaments, we utilized the SFUMATO code \citep{2007PASJ...59..905M}, a publicly available MHD simulation code
with the adaptive mesh refinement (AMR) method.
In our simulations, the equations for ideal MHD with self-gravity,
\begin{align}
\frac{\partial\rho}{\partial t}&+\nabla\cdot(\rho\boldsymbol{v})=0,\\
\frac{\partial}{\partial t}(\rho\boldsymbol{v})&+\nabla\cdot\left[\rho\boldsymbol{v}\boldsymbol{v}^T+\left(p+\frac{|\boldsymbol{B}|^2}{8\pi}\right)I-\frac{\boldsymbol{B}\boldsymbol{B}^T}{4\pi}\right]=-\rho \nabla \Psi ,\\
\frac{\partial\boldsymbol{B}}{\partial t}&=\nabla\times(\boldsymbol{v}\times\boldsymbol{B}),\\
\nabla^2 \Psi &=4 \pi G \rho,
\end{align}
are solved.
Here, $\rho$ is the gas density,
$\boldsymbol{v}$ is the gas velocity,
$\boldsymbol{B}$ is the magnetic field,
$\Psi$ is the gravitational potential,
$G$ is the gravitational constant,
$I$ is the unit matrix and superscript $T$ denotes the transposed matrix.
We assume the isothermal cloud  and the equation of state is given as,
\begin{equation}
P(\rho)={c_s}^2\rho,
\end{equation}
where $c_s=1.9\times 10^4 {\rm ~cm ~s^{-1}}$ is the sound velocity at $T \sim 10~\rm{K}$.

We consider the filament along the $x$-axis and the initial density profile of the filament is given by
\begin{equation}
\label{rho_prof}
\rho(R)=\rho_0\left(1+\frac{R^2}{{R_0}^2}\right)^{-2},
\end{equation}
where $R=\sqrt{y^2+z^2}$ is the cylindrical radius, $\rho_0=1.45\times 10^{-19}~\rm{g~cm^{-3}}$ is the density on the central axis of the filament ($x$-axis),
and $R_0=0.05$ pc is the scale height of the filament \citep{1964ApJ...140.1056O}. Note that $2 R_0=0.1$ pc is the typical width of the observed filament \citep{2010A&A...518L.102A}.

It should be highlighted that the observations suggest that the filament radial profile is proportional to $R^{-2}$ (e.g., \citet{2011A&A...529L...6A}, \citet{2016A&A...592A..54A}, \citet{2019A&A...623A..16S}) at large $R$ which is different from the density profile adopted in this paper and previous theoretical studies \citep{2015MNRAS.449L.123M, 2019ApJ...881...11M} ($\rho \propto R^{-4}$ at large $R$). Simulations with the initial conditions consistent with observed profiles are important and may change the filament evolution. However, in this study, the density profile of equation (\ref{rho_prof}) is employed to investigate filament evolution for consistency with previous theoretical studies.

We imposed an isotropic turbulent velocity field with an initial 3D root-mean-square Mach number of $\mathcal{M}=1$ and a Kolmogorov-like power spectrum of $P(k)\propto k^{-4}$.
This is consistent with previous simulations studies \citep{2015MNRAS.449L.123M, 2024ApJ...963..106M}
\footnote{Our initial condition for turbulence is consistent with the "1D Kolmogorov spectrum" of \citet{2019ApJ...881...11M}. 
We consider an isotropic 3D turbulence without cutoff in all axial directions so that the spectrum in each direction is given as,
\begin{equation}
    P(k)dk_xdk_ydk_z\propto k^{-4}k^2 dk=k^{-2}dk,
\end{equation}
which is roughly consistent with the Kolmogorov turbulence.
This also agrees with the 1D spectrum of \citet{2019ApJ...881...11M}.
On the other hand, the "3D Kolmogorov spectrum" of \citet{2019ApJ...881...11M} imposed different cutoff scales along the minor and major axial directions of the filament (i.e., their 3D turbulent field is anisotropic).
}.
We only impose the turbulence in the initial state and do not continuously drive the turbulence during the simulations.

These conditions are consistent with those of \citet{2015MNRAS.449L.123M}. 
In addition, we impose a magnetic field perpendicular to the filament, suggested by the observations of dust polarization (e.g., \citet{2019FrASS...6...15P}).
We initially imposed the constant magnetic field along the $y$-axis.
The initial strength of the magnetic field is the parameter and we adopt the values of $B_{\rm init}=7.23\times 10^{-1}~\rm{\mu G}$, $2.16~\rm{\mu G}$, and $7.23~\rm{\mu G}$.

The magnetic flux within the cylindrical radius of $R_0 =0.05\ \rm{pc}$ is calculated as
\begin{equation}
    \Phi=\lambda \int_{R=-R_0}^{R=R_0} B  dR=2 \lambda R_0 B, 
\end{equation}
where $\lambda$ is the length along the filament. We also assume a constant magnetic field.
Next, the mass within the same region is calculated as
\begin{equation}
    \label{eq_mass}
    M=\lambda \int_0^{R=R_0} {2\pi R \rho(R) }dR.
\end{equation}
Using these estimates, the mass-to-flux ratio of a filament, as a function of the magnetic field strength, is given by
\begin{equation}
    \mu=\frac{(M/\Phi)}{(M/\Phi)_{\rm{crit}}}=18.87 \left(\frac{B}{\mu \rm{G}}\right)^{-1}.
\end{equation}
 Thus, the initial mass-to-flux ratio of our filament model is $\mu=26-2.6$, assuming the critical mass-to-flux ratio for filaments of $(M/\Phi)_{\rm{crit}}=0.24/\sqrt{G}$ \citep{2014ApJ...785...24T}. 

 Note that the magnetic field strength employed in this paper is relatively weak compared to the values obtained from observations (e.g. \citet{2022MNRAS.510.6085L}) which implies the filament is supercritical. 
Stronger magnetic fields are expected to play a broader role, which will be discussed in more detail in section \ref{sec_conclusion}.

The simulation domain is a cubic box of $1.5~\rm{pc}$ per side and is resolved on a base grid with $128^3$ cells.
Periodic boundary conditions were applied.
Grid blocks are refined by requiring the Jeans length to be resolved with at least 8 grids.
The highest refinement level from the base grid ($l=0$) is $l=4$ and the finest cells have the resolution of $\Delta x=7.3 \times 10^{-4}\ \rm{pc}$.

The parameters of the initial filament are summarized in Table \ref{tab_initcondi} (for the configuration of the simulation, see also figure \ref{fig_def_angle}).

\begin{table*}[h]
 \caption{Initial condition of 3D-MHD numerical simulations.}
 \label{tab_initcondi}
 \centering
  \begin{tabular}{llcl}
   \hline\hline
   \multirow{7}{*}{Physical state of filament}
   & Cylindrical radius & $R_0$ & $0.05~\rm{[pc]}$\\
   & Length & $L$ & $1.5~\rm{[pc]}$\\
   & Density at central axis & $\rho_0$ & $1.45 \times10^{-19}~\rm{[g~cm^{-3}]}$\\
   & Density radial profile & & $\rho(R)=\rho_0{\left(1+\frac{R^2}{{R_0}^2}\right)}^{-2}$\\
   & line mass\footnotemark[$*$] & $M_{\rm line}$ & $16.6~\rm{[M_{\odot}~pc^{-1}]}$ \\
   & Jean's length & & $\lambda_J=c_s \left(\frac{\pi}{G\rho_0}\right)^{1/2}=3.42\times 10^{18} ~\rm{[cm]}\simeq 1.1~\rm{[pc]}$\\
   & Free-fall time & & $t_{\rm{ff}}=\left(\frac{3\pi}{32G\rho_0}\right)^{1/2}\sim1.7\times 10^{5}~\rm{[yr]}$\\
   \hline
   Turbulence & 3D rms Mach number & & $\mathcal{M}=1$\\
   \hline
   \multirow{3}{*}{\shortstack{Magnetic field~(only in $y$-direction)\\(Normalized mass-to-flux ratio)}}
   & Very weak & & $7.23\times 10^{-1}~\rm{[\mu G]}~(\mu_0=26.1)$\\
   & Weak & & $2.16~\rm{[\mu G]}~(\mu_0=8.74)$\\
   & Moderate & & $7.23~\rm{[\mu G]}~(\mu_0=2.61)$\\
   \hline
  \end{tabular}
  \begin{tablenotes}
    \item \footnotemark[$*$] $M_{\rm line}$ was derived as $R\rightarrow\infty$ in equation (\ref{eq_mass})
  \end{tablenotes}
\end{table*}

\subsection{Synthetic observation}
We used RADMC-3D to calculate polarized dust emission at a wavelength of $\lambda=850~\rm{\mu m}$ which is the observational wavelength of the BISTRO survey.
The dust and gas temperature is assumed to be $T=10~\rm{K}$.
Oblate dust grains with a size of $1~\rm{\mu m}$ and composed of amorphous pyroxene $\rm{Mg(0.7)Fe(0.3)SiO(3)}$ are considered.
The optical constants of dust grains are taken from \citet{1994A&A...292..641J}.
The dust density distribution is calculated from the simulated gas density distribution by assuming a dust-to-gas ratio of $\rho_{\rm dust}/\rho_{\rm gas}= 0.01$.

In RADMC-3D, the alignment of dust grains is determined by the alignment vector field $\boldsymbol{p}_{\rm{align}}=(p_x, p_y, p_z)$ at each cell of the radiative transfer simulations.
Here, we assume,
\begin{equation}
  (p_x,~p_y,~p_z)=\left( \frac{\epsilon_{\rm{align}}B_x}{|\boldsymbol{B}|},\frac{\epsilon_{\rm{align}}B_y}{|\boldsymbol{B}|},\frac{\epsilon_{\rm{align}}B_z}{|\boldsymbol{B}|}\right).
  \label{eq_alignment_vector}
\end{equation}
The alignment degree $\epsilon_{\rm{align}}$ is set to be a constant value of $0.1$ for simplicity,
and we focus on the impact of the magnetohydrodynamical effect on the polarized emission.
More realistically, the spatial variation of the alignment degree should be taken into account, for example, considering the Radiative Torque Theory (RAT) \citep{2008MNRAS.388..117H}.

\begin{figure}
  \begin{center}
      \includegraphics[width=80mm]{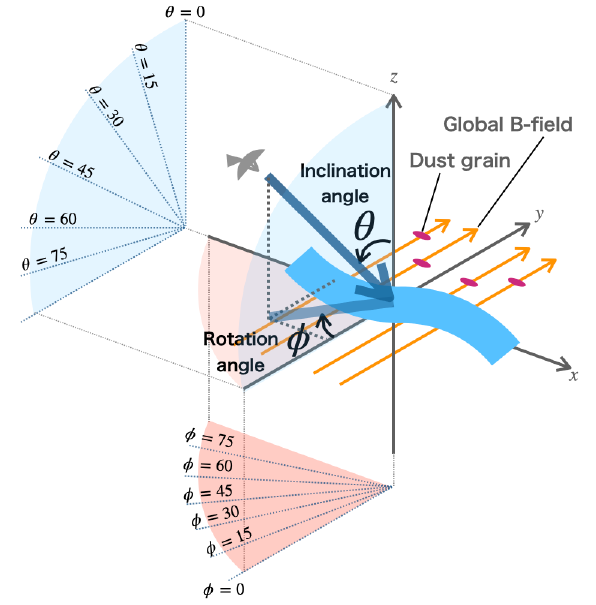}
  \end{center}
  \caption{The definition of the angles to specify the line-of-sight(LOS).
    The inclination angle $\theta$ is the angle between the LOS and the $z$-axis. The rotation angle $\phi$ is the angle between the $y$-axis and the projected LOS on the $x-y$ plane. 
    The LOS is perpendicular to the initial magnetic field
    when $\theta = 0^\circ$ and is parallel to the initial magnetic field when $(\theta, \phi) = (90^\circ, 0^\circ)$.
    The major axis of the filament is on the plane-of-sky (POS) when $\phi = 0^\circ$.}
  
  \label{fig_def_angle}
\end{figure}

The polarized intensity also depends on the line-of-sight (LOS) direction.
We define the two angles to specify the LOS: the inclination angle $\theta$ which is the angle between the LOS and the $z$-axis,
and the rotation angle $\phi$ which is the angle between the projected LOS onto $x$-$y$ plane and the $y$-axis (see figure \ref{fig_def_angle}).
We mainly examine the polarized dust emission from the filament at $t \sim 2 t_{\rm{ff}}$.

\section{Results}
\label{sec_results}
We describe the filament evolution obtained from 3D-MHD simulations in section \ref{subsec_sim_results}. We then examine the synthetic observations from the MHD simulations in section \ref{subsec_synob_results}.
In section \ref{subsec_relation_bw_BandP}, we discuss the origins of the polarization profiles obtained in section \ref{subsec_synob_results}.

The simulation with $B_{\rm init}=\mb$ is adopted as the fiducial model and is mainly discussed in this section.

\subsection{Filament fragmentation and cloud core evolution}
\label{subsec_sim_results}

\begin{figure*}
\begin{center}
\includegraphics[width=170mm]{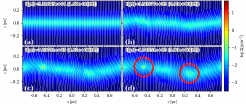}
\end{center}
\caption{Time evolution of the surface density on the $x$-$y$ plane of the filament with magnetic field strength of $|\boldsymbol{B}_{\rm init}| = 2.16\ \mu\mathrm{G}$,
  at the epoch of $t \approx 0.2$, $2$, $4$, and $5 \times t_{\rm{ff}}$ (free-fall time), from (a) to (d).
  The two dense cores indicated by the red circles are formed in panel (d). The white lines show the density-weighted magnetic field lines.}
\label{fig_sim_filament}
\end{figure*}

Figure \ref{fig_sim_filament} shows a column density map on the $x$-$y$ plane for the model with $B_{\rm init}=\mb$ at $t\sim 0.2,~2,~4,~5\times t_{\rm{ff}}$, where $t_{\rm{ff}}$ is calculated from the initial density on the central axis of the filament (see Table \ref{tab_initcondi}).

\begin{figure}
  \begin{center}
      \includegraphics[width=90mm]{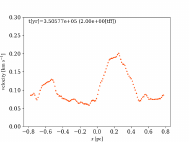}
  \end{center}
  \caption{Density weighted mean velocity along the filament at $t=2\times t_{\rm ff}$.The vertical axis shows the mean velocity calculated by equation (\ref{eq_meanvel}) on the $y-z$ plane at position $x$.}
  \label{fig_velocity_dispersion}
\end{figure}

The figure shows that the filament fragments into two cloud cores (indicated by red circles) on a timescale of several free-fall times.
We confirmed that the fragmentation happened regardless of the initial magnetic field strength considered in this paper.
The separation between the two cores is approximately $0.6~\rm{pc}$, slightly wider than the prediction of linear perturbation theory
($\sim 4\times R_0 = 0.4~\rm{pc}$) \citep{1997ApJ...480..681I}, mainly due to additional pressure support by turbulence.

Figure \ref{fig_velocity_dispersion} shows the mean velocity in the filament at $t\sim 2\times t_{ff}$ (corresponding to the epoch of figure \ref{fig_sim_filament} (b)). 
The mean velocity at position $x$ is calculated as
\begin{equation}
    \bar{v}(x)=\frac{\int (\rho | \mathbf{v}| )~ dy dz }{\int \rho~ dy dz},
    \label{eq_meanvel}
\end{equation}
where the integration was performed on the y-z plane.
The figure shows that the velocity field in the filament is subsonic and consistent with the observations (e.g. \citet{2013A&A...553A.119A}).

Because both observational and theoretical studies suggested the increase in specific angular momentum of the cores
as they grow \citep{1993ApJ...406..528G, 2016PASJ...68...24T,2019ApJ...881...11M}, we investigated the specific angular momentum evolution of cores formed in our MHD simulations.
We defined a cloud core as a region with a density more than 100 times larger than the initial density of the central axis, $\rho_0$.
Hence the core mass is given as,
\begin{equation}
    M_{\rm{core}}=\int_{\rho>100\rho_0}\rho dV.
\end{equation}
Then we define the position and velocity of the core as, 
\begin{align}
    \boldsymbol{r}_G=\frac{\int_{\rho>100\rho_0} \rho \boldsymbol{r}dV}{M_{\rm{core}}},\\
    \boldsymbol{v}_G=\frac{\int_{\rho>100\rho_0} \rho \boldsymbol{v}dV}{M_{\rm{core}}}.
\end{align}
The angular momentum and specific angular momentum of the core are given as
\begin{align}
    \boldsymbol{J}_{\rm{core}}&=\int \rho \boldsymbol{r}'\times \boldsymbol{v}' dV',\\
    \boldsymbol{j}&=\frac{\boldsymbol{J}_{\rm{core}}}{M_{\rm{core}}},    
\end{align}
respectively, where 
\begin{align}
    \boldsymbol{r}'=\boldsymbol{r}-\boldsymbol{r}_{G},\\
    \boldsymbol{v}'=\boldsymbol{v}-\boldsymbol{v}_{G},    
\end{align}
are the relative position and velocity.

\begin{figure}
\begin{center}
\includegraphics[width=90mm]{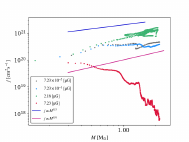}
\end{center}
\caption{
    Time evolution of specific angular momentum as a function of core mass. This figure can be directly compared with figure 1 of \citet{2019ApJ...881...11M}.
    The dots show the data from the core first formed in our simulations. The solid lines indicate the empirical power laws suggested from observations; $j\propto M^{0.5}$ \citep{1993ApJ...406..528G,2016PASJ...68...24T} and $j\propto M^{0.9}$ \citep{2002ApJ...565..331C}.
    The line colors represent the initial magnetic field strengths; light blue for $7.23 \times 10^{-1}\ \mu\mathrm{G}$, green for $2.16\ \mu\mathrm{G}$, red for $7.23\ \mu\mathrm{G}$, and gray for $0\ \mu\mathrm{G}$.}
\label{fig_jspe}
\end{figure}

Figure \ref{fig_jspe} shows the time evolution of specific angular momentum as a function of core mass.
Notably, the evolution of the specific angular momentum varies depending on the magnetic field strength.
For a weak magnetic field ($B \lesssim 2 {\rm \mu G}$) or without a magnetic field, the specific angular momentum evolution is consistent with the relation suggested by the observations.
In contrast, in the case of a moderate magnetic field ($B \sim 7 {\rm \mu G}$), the angular momentum significantly decreases with mass growth.
This suggests that there is a critical initial field strength of $B \sim 10 {\rm \mu G}$, at which the magnetic braking begins
to work for the evolution of the cores formed from the quasi-steady-state filament.

\subsection{Synthetic observations and the application of the polarization data analysis method proposed by \citet{2021ApJ...923L...9D}}
\label{subsec_synob_results}
\citet{2021ApJ...923L...9D} introduced a scalar quantity, $\rfwhm$, which characterizes the polarized dust emission as the ratio of the Full Width at Half Maximum (FWHM) of the polarized intensity ($PI$) to that of the total intensity ($I$).
They proposed that the three-dimensional magnetic field structure inside the filament could be revealed by analyzing this quantity.
In this subsection, we apply their methods to our turbulent magnetized filaments and discuss what information about the magnetic field can be obtained from the ratio of $FWHM$ of $PI$ to that of $I$.

\begin{figure*}
\begin{center}
\includegraphics[width=100mm]{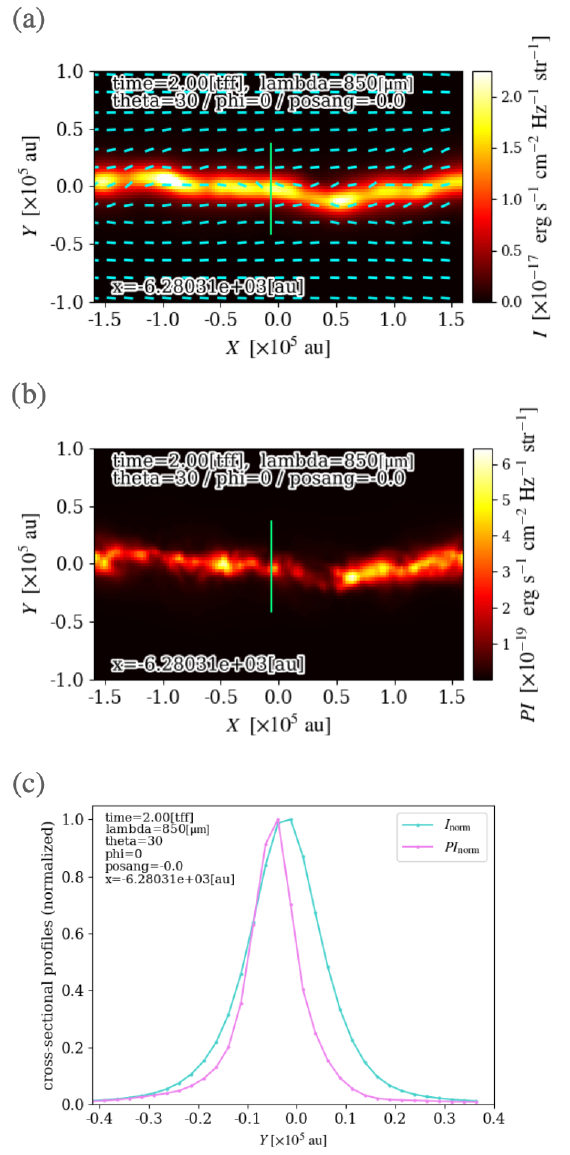}
\end{center}
\caption{
    The procedure of analysis employed in this study is shown using the case with LOS of $(\theta,~\phi)=(30^\circ,~0^\circ)$ as an example of synthetic observation results. 
    Panel (a) shows the total intensity ($I$) map and polarization direction (blue lines).
    Panel (b) shows the polarized intensity ($PI$) map.
    Panel (c) shows the profiles of $I$ (blue line) and $PI$ (pink line) along the green line shown in panels (a) and (b), respectively.
    The profiles are normalized by their maximum values.}
\label{fig_synob_overview}
\end{figure*}

We provide our analysis procedure in figure \ref{fig_synob_overview}.
First, we compute 2D maps of $I$ and $PI$ from various LOS using RADMC-3D
(an example with LOS with $\theta=30$ and $\phi=0$ is shown in panels (a) and (b), and for the definition of $\theta$ and $\phi$, see figure \ref{fig_def_angle}.).
We then plot $I$ and $PI$ along the minor axis of filament (indicated by the green line in panels (a) and (b)), which is perpendicular to the $x$-axis projected onto the POS (see figure \ref{fig_def_angle}).
Panel (c) shows the profiles of $I$ and $PI$ along the green line.

A Gaussian fitting is then applied to the $I$ and $PI$ profiles along the minor axis using the nonlinear least squares method.
The Gaussian function is represented as follows:
\begin{equation}
f(x)_{\mu,\sigma}=\frac{1}{\sqrt{2\pi\sigma^2}}\exp{\left\{-\frac{(x-\mu)^2}{2\sigma^2}\right\}},
\end{equation}
where $\mu$ and $\sigma$ denote the mean and standard deviation, respectively,
and the $FWHM$ is calculated using the following formula:
\begin{equation}
FWHM = 2\sigma\sqrt{2\ln{2}}.
\end{equation}
Then, the ratio of the derived $FWHM$ values of $PI$ to $I$ is calculated as
\begin{equation}
\rfwhm=\frac{FWHM_{\rm PI}}{FWHM_{\rm I}}.
\end{equation}

We calculate $\rfwhm$ on 128 points (the number of the base grid per side)
along the $X$-axis (by changing the position of the green line) for various $\theta$ and $\phi$,
where the $X$-axis is the direction of the major axis projected on the plane-of-sky (POS).

\begin{figure*}
\begin{center}
\includegraphics[width=170mm]{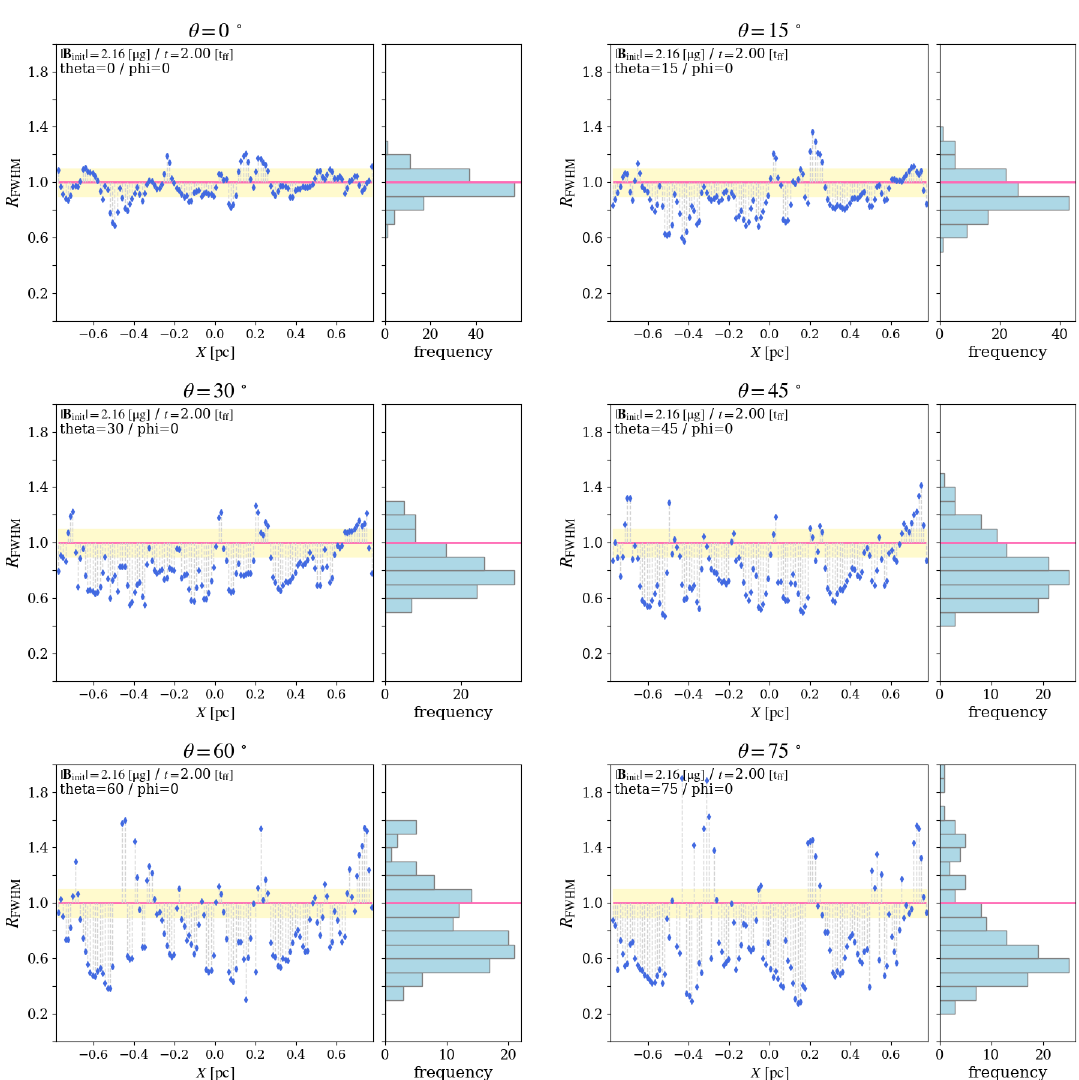}
\end{center}
\caption{
    $\theta$ dependency of the distribution of $\rfwhm$ within a filament. The left side of each panel displays the $\rfwhm$.
    The vertical and horizontal axes represent the $\rfwhm$ and the position along the filament major axis, respectively.
    The pink line marks where $\rfwhm = 1.0$ and the region shaded in yellow represents the range $0.9 \leq \rfwhm \leq 1.1$.
    The right side of each panel shows histograms of the $\rfwhm$ distribution presented in the left side panel. In these plots, the rotation angle is set to $\phi=0^\circ$.}
\label{fig_scat_histo}
\end{figure*}

Figure \ref{fig_scat_histo} shows the $\rfwhm$ distribution along the major axis of the filament for various $\theta$ with $\phi=0$.
The left panels show the distribution of $\rfwhm$ along the $X$-axis, while the right panels display the corresponding histogram.
We can see two interesting properties from the histogram of figure \ref{fig_scat_histo}.
First, the histogram peak shifts toward a smaller value of $\rfwhm$ with increasing $\theta$.
Second, the variance of $\rfwhm$ increases with increasing $\theta$.
These two properties suggest that the distribution of $\rfwhm$ is related
to the inclination of the global magnetic field to the POS, which is difficult to measure.

To evaluate quantitatively these features, we calculate the mean $\rbar$ and variance $\rsigma$ of $\rfwhm$,
\begin{align}
\label{eq_mean} \rbar&=\frac{\sum\limits_{n=1}^{N} \rfwhm}{N},\\
\label{eq_variance} \rsigma&=\frac{\sum\limits_{n=1}^{N} (\rfwhm-\rbar)^2}{N},
\end{align}
where $N=128$ is the number of data samples, corresponding to the number of grids along the $X$-axis.

\begin{figure*}
\begin{center}
\includegraphics[width=170mm]{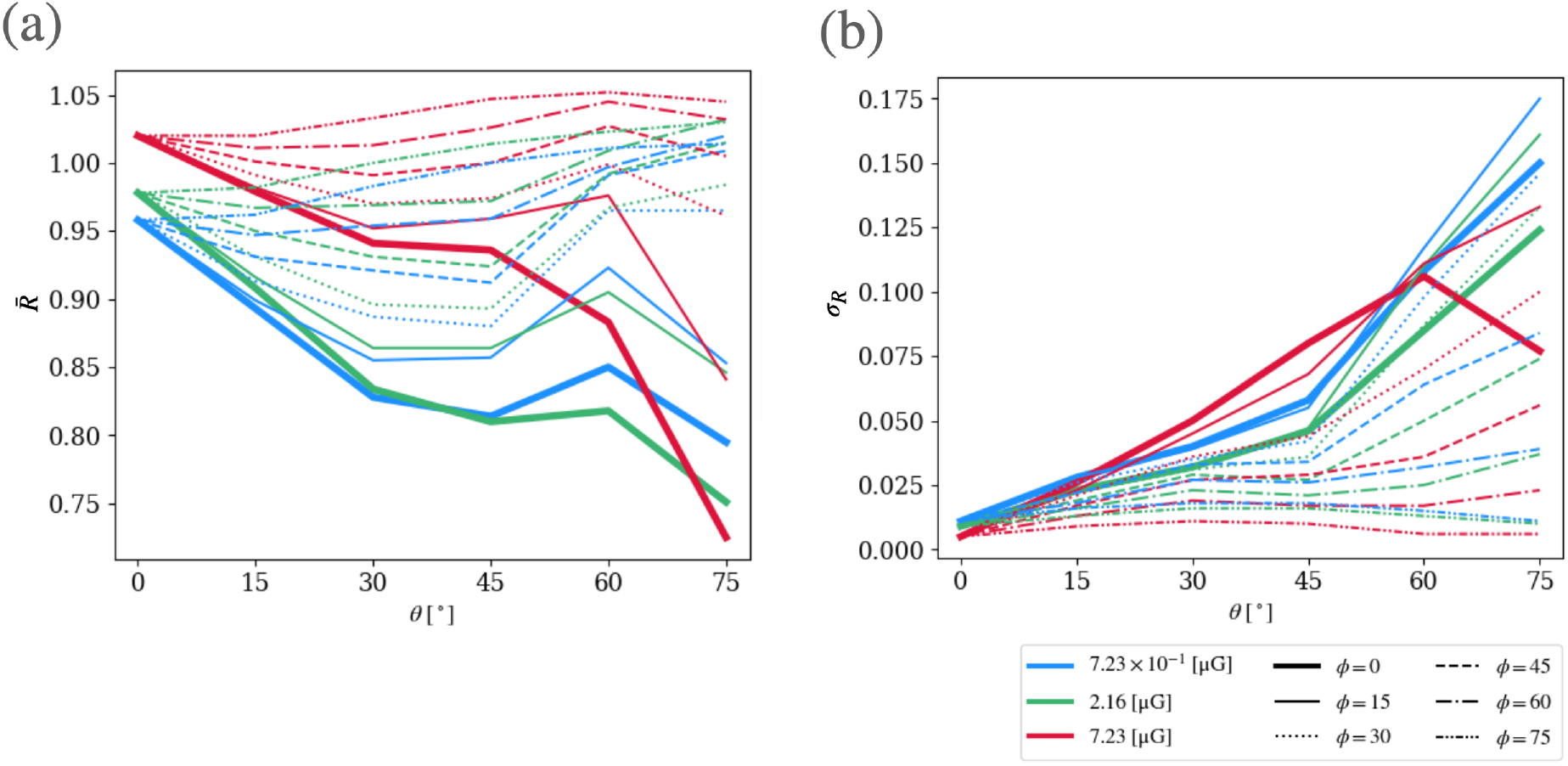}
\end{center}
\caption{(a) Mean ($\rbar$) and (b) variance ($\rsigma$) of $\rfwhm$ within the filament at the epoch of $t\sim2\times t_{\rm ff}$ as a function of $\theta$.
  The line colors indicate the initial magnetic field strengths;
  light blue for very weak field ($\wb$), green for weak field ($\mb$), and red for moderate field ($\stb$).
  The line styles represent the rotation angle ($\phi$).}
\label{fig_barsigma_thetaphib_dep}
\end{figure*}

Figure \ref{fig_barsigma_thetaphib_dep} shows (a) $\rbar$ and (b) $\rsigma$ as a function of $\theta$.
The line styles represent the different values of $\phi$ and the line colors correspond to the initial magnetic field strength.
The three thick solid lines clearly show that $\rbar$ decreases and $\rsigma$ increases with increasing $\theta$.
The lines with different field strengths (different colors) show the same trend. 
This suggests that the trend is independent of the initial magnetic field strength.
However, comparing the different line styles of the same color (representing different values of $\phi$) reveals that the trend becomes weaker as $\phi$ increases.
For example, the decrease in $\rbar$ seen in the solid line in panel (a) almost disappears once $\phi \gtrsim 45^\circ$. 
The same is true for $\rsigma$, where the increase in it almost disappears for $\phi \gtrsim 45^\circ$.

This result suggests that the inclination of the global magnetic field of a filament relative to the POS can be constrained by analyzing the $\rfwhm$ distribution within the filament.
For example, observational values of $\rbar < 0.9$ and $\rsigma > 0.1$ indicate that the inclination angle $\theta \gtrsim 60^\circ$ and the rotation angle $\phi \lesssim 15^\circ$, as shown in figure \ref{fig_barsigma_thetaphib_dep}.

This reminds us of the large scatter in $\rfwhm$ values reported by \citet{2021ApJ...923L...9D}, where $\rfwhm$ varies widely, ranging from a minimum of $\sim 0.8$ to a maximum of $\sim 1.3$ across the four observed segments.
This result implies that the magnetic field in the filament they observed is likely to be strongly inclined relative to the POS.

\subsection{Relation between $\rfwhm$ and disordered magnetic field structure inside the filament}
\label{subsec_relation_bw_BandP}

In this section, we explore why $\rfwhm$ varies within a filament
and examine the trends of $\rbar$ decreasing and $\rsigma$ increasing with increasing $\theta$ in turbulent filamentary molecular clouds.
In the following discussion in this section, 
we focus on the case of $\phi=0$, which means the LOS is always perpendicular to the major axis of the filament.

A possible mechanism by which turbulence can reduce the polarized intensity is that the polarization components cancel each other out due to the fluctuations of the magnetic field along the LOS.
Such magnetic field fluctuations can be caused by turbulent eddies within the filament (i.e., eddies with the size of $<0.1\ \rm pc$). 
We call this effect process (I).

On the other hand, \citet{2021ApJ...923L...9D} pointed out that when the hourglass magnetic field configuration
is viewed from an oblique angle, the magnetic field in the filament outer region tends to parallel with the LOS,
resulting in weaker polarized emission from the outer region toward the LOS.
A similar effect can be caused by turbulent eddies with the scale of filament width (i.e., eddies with the size of $\gtrsim 0.1\ \rm pc$). 
Such filament-scale eddies can tilt the magnetic field toward the LOS.
We call this effect of large-scale ($\gtrsim 0.1 \ \rm pc$) turbulent eddies that tilt the magnetic field toward the LOS and reduce the polarized intensity process (II).

Note that the filament considered in this study is qualitatively different from the filament model considered in \citet{2021ApJ...923L...9D}.
They considered a filament in which magnetic pressure and tension mainly balance the self-gravity (i.e., the filament is massive so that  the gas pressure cannot balance the self-gravity), where the hourglass morphology arises naturally.
On the other hand, in the filament investigated in this study, the thermal pressure mainly balances the self-gravity. Hence, we do not expect the hourglass morphology in the filament. Indeed, we did not find the hourglass morphology of the magnetic field on the $y$-$z$ plane, except inside the collapsing cloud cores, where the magnetic field is bent toward the center. 
Therefore, in the filament we assume, the main physical process that tilts the magnetic field toward the LOS would be the turbulent eddies with the scale of the filament width ($\gtrsim 0.1\ \rm{pc}$).

To investigate the above two processes separately, we examined the distribution of the following two quantities in cross-sections of the filament (i.e., on the $y$-$z$ plane; see figure \ref{fig_def_angle}).

\begin{figure*}
 \begin{center}
    \includegraphics[width=170mm]{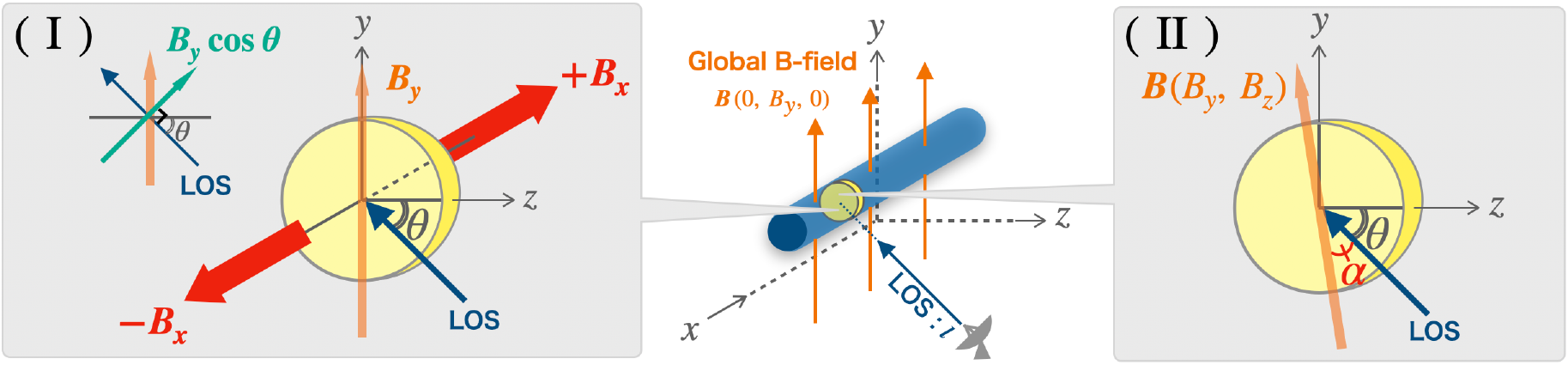}
 \end{center}
 \caption{
   Schematic figure of our analysis to investigate the mechanism of depolarization in a turbulent filament.
   We investigate two depolarization processes on a cross-section of filament; (I) the depolarization due to the magnetic
   field fluctuation along the LOS and (II) the alignment of the magnetic field toward the LOS.
   Panel (I) shows the direction of the $B_x$ component which is perpendicular to both the LOS and the $y$-axis.
   By investigating the fluctuation of $B_x$ along the LOS, we can evaluate depolarization along the LOS. 
   Panel (II) shows the definition of the angle $\alpha$ which represents the degree of the alignment between the LOS and the magnetic field.}
  \label{fig_illustrate_analysis}
\end{figure*}

For process (I), we examined the following quantity,
\begin{equation}
    \chi_I = \frac{\rho}{\rho_{\rm{max}}}\times\frac{B_x}{B_y\cos{\theta}}.
    \label{eq_factor1}
\end{equation}
$B_x$ is the magnetic field component, which is perpendicular to both the LOS and the initial magnetic field (i.e., magnetic field along filament major axis),
and $B_y\cos{\theta}$ is the magnetic field component perpendicular to the LOS (see figure \ref{fig_illustrate_analysis}~(I) for the geometry of each component). 

\begin{figure}
\begin{center}
\includegraphics[width=80mm]{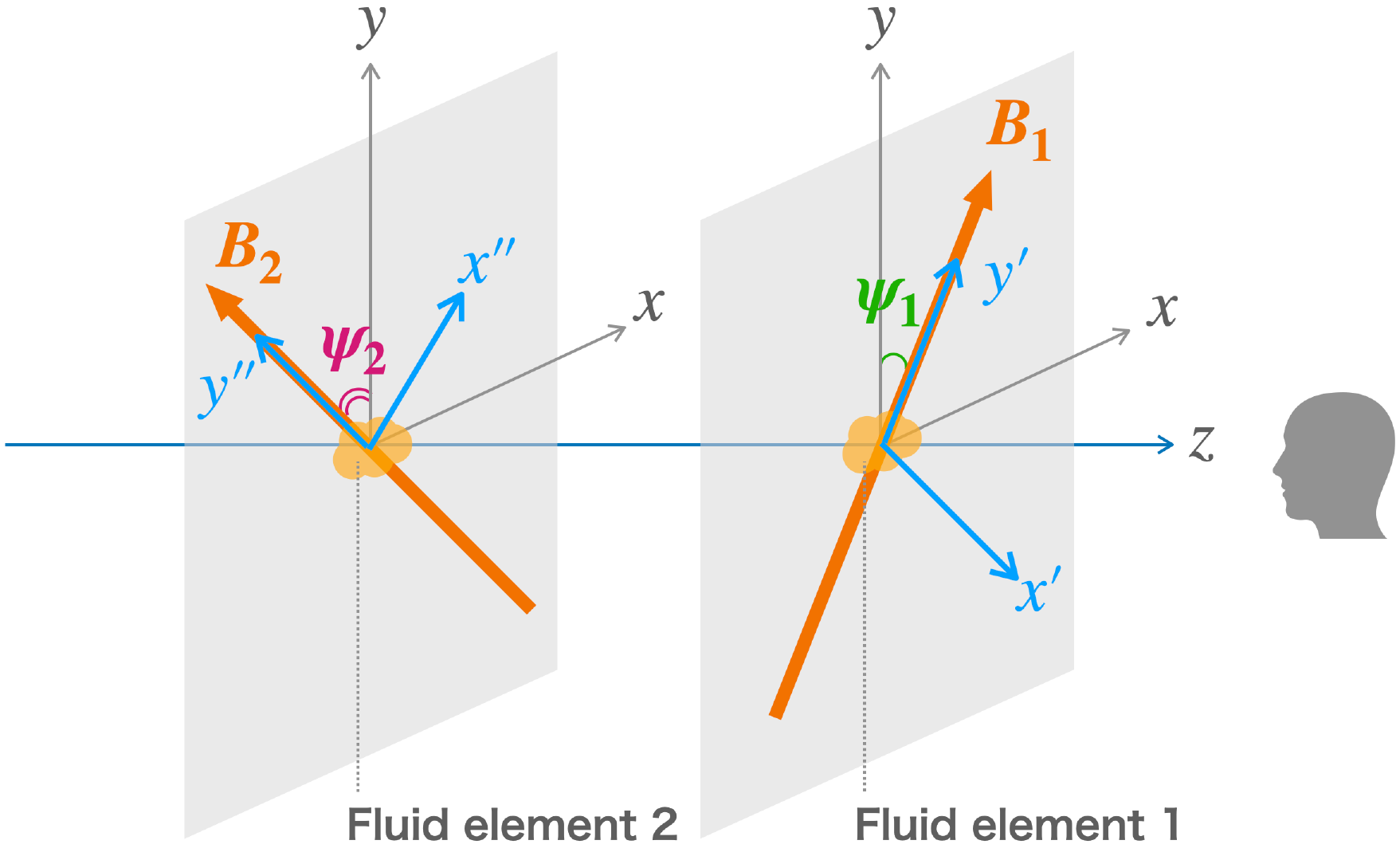}
\end{center}
\caption{ Schematic figure to explain the meaning of $\chi_I$ (equation (\ref{eq_factor1})). 
          The coordinate system corresponds to the case of $\theta=0$ in figure \ref{fig_illustrate_analysis}, where the LOS coincides with the $z$-axis direction. 
          The $x$-axis and $y$-axis correspond to the horizontal and vertical axes on the POS, respectively. 
          $\psi$ is the offset angle of the magnetic field projected onto the POS from the $y$-axis of the fluid element. 
          We consider the effect on the $PI$ when the two fluid elements with different $\psi$ values are aligned along the LOS.}
\label{fig_illustrate_chi1}
\end{figure}

The meaning of $\chi_I$ can be understood by considering the superposition of the polarized intensity emitted from two (identical) radiative fluid elements aligned on the LOS.
Using the schematic figure shown in figure \ref{fig_illustrate_chi1}, this is illustrated as follows.
The $z$-axis is along the LOS, and the $y'$- and $y''$-axis correspond to the direction of the projected magnetic field on the POS at the position of two fluid elements, respectively. 
Then, we define $dI_{x'}$ and $dI_{y'}$, $dI_{x''}$ and $dI_{y''}$ as the contributions to the intensities of light polarized along the $x'$- and $y'$-, $x''$- and $y''$- axes. 
Note that $dI_{x'}=dI_{x''}$, $dI_{y'}=dI_{y''}$ because we assume two identical fluid elements.

Then the Stokes parameters $U$ and $Q$ in the fixed $x$-$y$ coordinate system from a radiative element are given as
\begin{equation}
  \begin{pmatrix}
    dQ \\
    dU
  \end{pmatrix}
  =
  \begin{pmatrix}
    \cos{(2\psi)}(dI_{x'}-dI_{y'}) \\
    \sin{(2\psi)}(dI_{x'}-dI_{y'})
  \end{pmatrix},
\end{equation}
where $\psi$ is the angle between the $y$-axis and the projected magnetic field onto the $x$-$y$ plane \citep{1990ApJ...362..120W}.
When each element has $\psi_1$ and $\psi_2$, respectively,  the sum of $dQ$ and $dU$ by the two elements is given as,
\begin{equation}
  \begin{pmatrix}
    dQ \\
    dU
  \end{pmatrix}
  =
  \begin{pmatrix}
    \cos{(2\psi_1)}+\cos{(2\psi_2)} \\
    \sin{(2\psi_1)}+\sin{(2\psi_2)}
  \end{pmatrix}
    (dI_{x'}-dI_{y'}).
\end{equation}
Then the integrated polarized intensity from the two elements is given as,
\begin{equation}
  \begin{aligned}
    PI
    &=\sqrt{(dQ)^2+(dU)^2}\\
    &=|2\cos{(\delta\psi)}(dI_{x'}-dI_{y'})|,
  \end{aligned}
\end{equation}
where $\delta\psi=(\psi_2-\psi_1)$.
When $\delta\psi=\pi/2$, the polarized emission from the two elements is completely depolarized (i.e., $PI=0$).
$B_x/(B_y\cos{\theta})$ in the right-hand side of equation (\ref{eq_factor1}) corresponds to $\tan{\psi}$.
Thus, if regions with positive and negative $\chi_I$ exist along the LOS, they reduce $PI$.

For process (II), we examined how well the LOS ($\boldsymbol{l}$)
and the magnetic field ($\boldsymbol{B}$) align using following quantity:
\begin{equation}
    \chi_{II} = \frac{\rho}{\rho_{\rm{max}}}|\sin{\alpha}|,
    \label{eq_factor2}
\end{equation}
where $\alpha$ is the angle between $\boldsymbol{l}$ and $\boldsymbol{B}$ (see, figure \ref{fig_illustrate_analysis}(II)).
The closer $\boldsymbol{l}$ and $\boldsymbol{B}$ are to being parallel ($\alpha \rightarrow 0$),
the smaller the value of $\chi_{II}$ becomes, and vice versa.

\begin{figure*}
 \begin{center}
  \includegraphics[width=170mm]{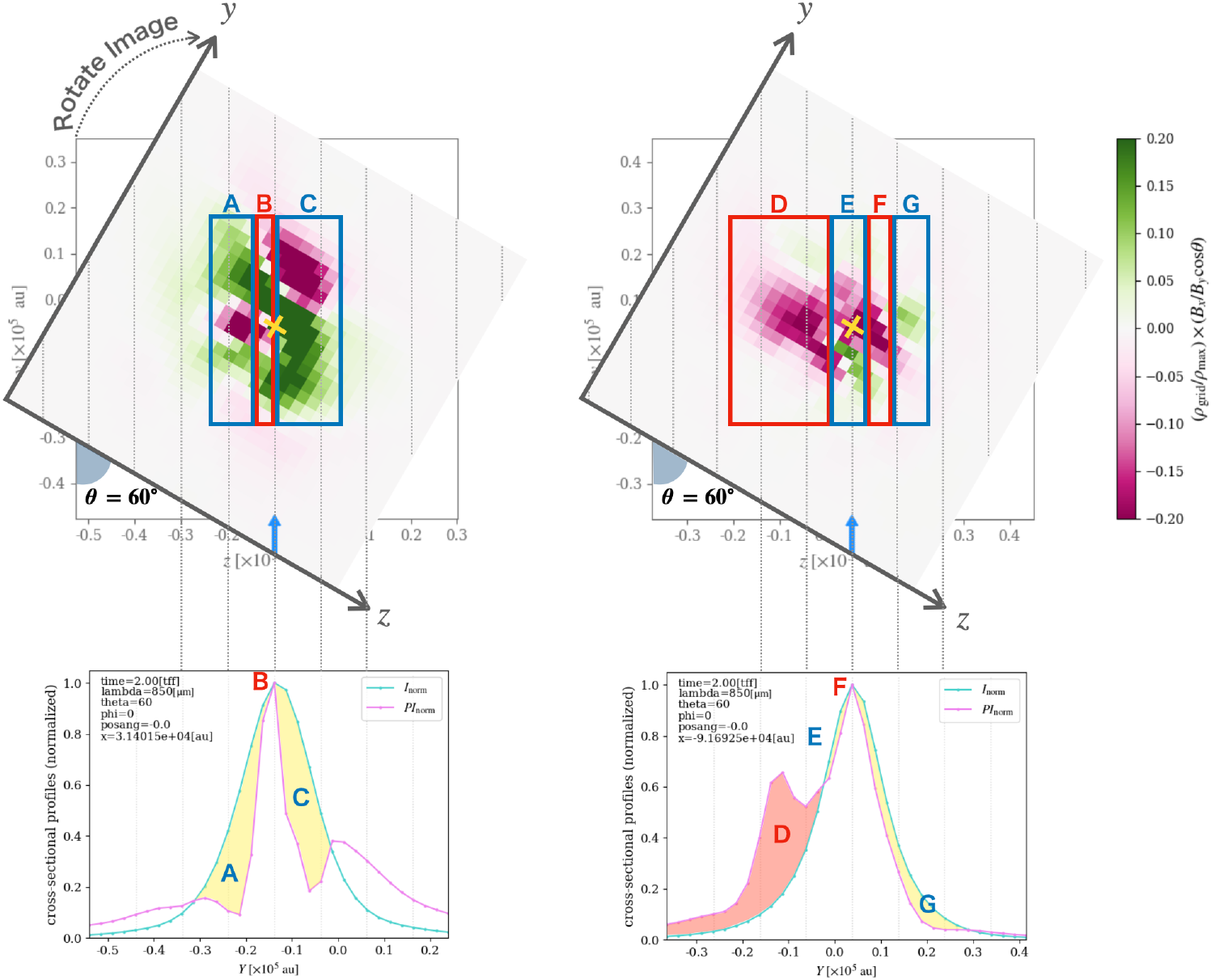}
 \end{center}
 \caption{
      The upper panels show the map of $\chi_I$ (equation (\ref{eq_factor1})) on a cross-section.      The panels correspond to the plane with the (a) minimum and the (b) maximum $\rfwhm$ in the filament, at $t=2 t_{\rm{ff}}$ with a LOS of $(\theta,~\phi)=(60^\circ,~0^\circ)$.
      The LOS is indicated with a blue arrow. 
      The gray dashed lines are parallel to LOS and drawn with the separations of $10000~\rm{au}$ from the density peak (indicated with a yellow cross).
      The lower panels show the corresponding polarization profile plotted in the same way as figure \ref{fig_synob_overview} (c).}
 \label{fig_discussion_I}
\end{figure*}

\begin{figure*}
 \begin{center}
   \includegraphics[width=160mm]{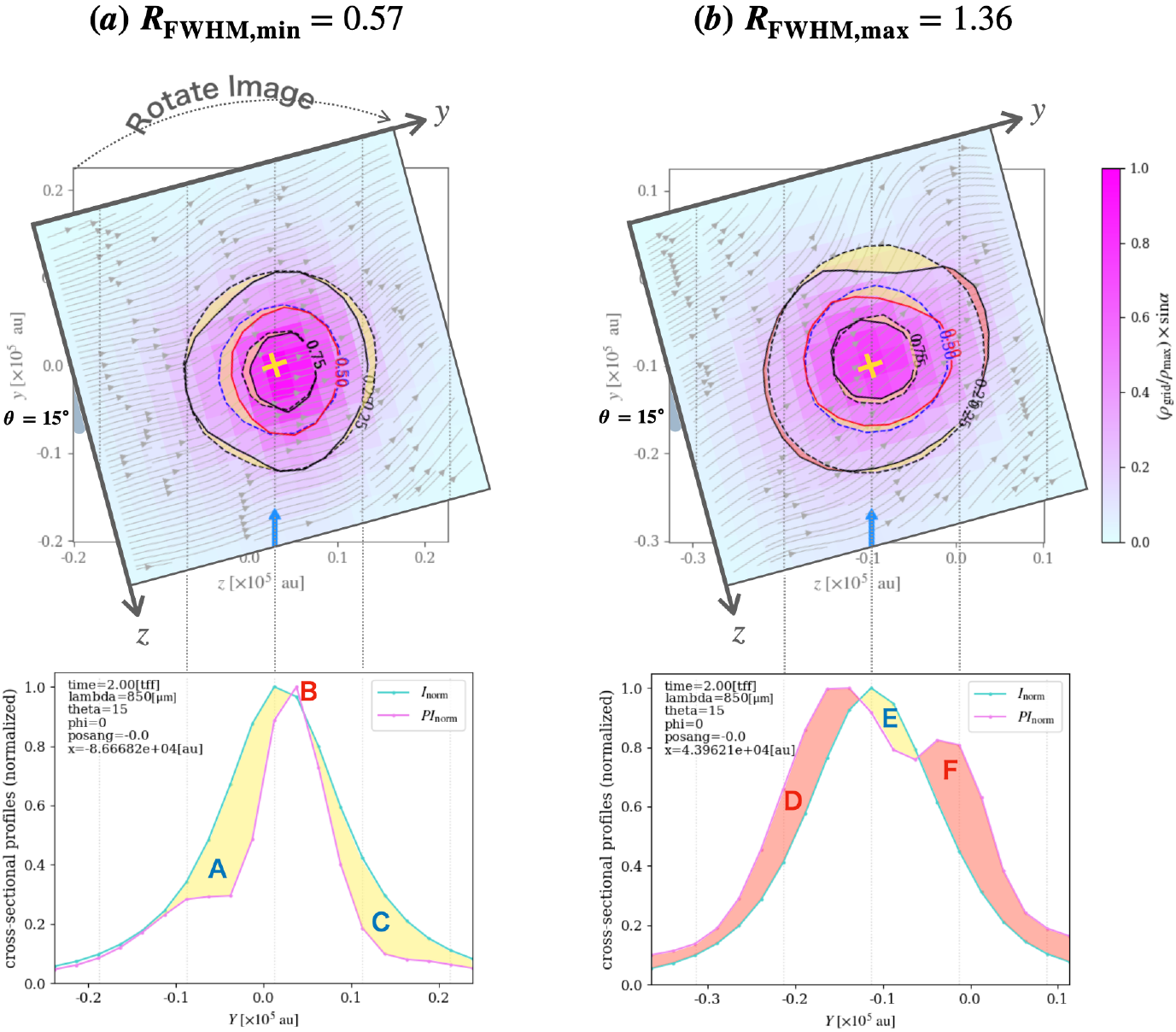} 
 \end{center}
 \caption{
    The upper panels display the map of $\chi_{II}$ (equation  (\ref{eq_factor2})) with its contours (solid lines) overlaid, as well as the density contours (dashed lines).
    Both contours are normalized by their maximum values. 
    The panels correspond to the plane with the (a) minimum and the (b) maximum $\rfwhm$ in the filament, at $t=2 t_{\rm{ff}}$ with a LOS of $(\theta,~\phi)=(15^\circ,~0^\circ)$.
    Yellow-filled areas in upper panels indicate where the normalized $\chi_{II}$ value is smaller than the normalized density, and red-filled areas where it is larger. 
    The light gray streamlines are the magnetic field at this plane.
    The lower panels show the corresponding polarization profile plotted in the same way as figure \ref{fig_synob_overview} (c), but the plot range of the horizontal axis is extracted near the peak position.}
 \label{fig_discussion_II}
\end{figure*}

We examined $\chi_I$ and $\chi_{II}$ across almost all the cross-sections of the simulation results.
Among them, we show typical examples in which $PI$ is significantly affected by processes (I) and (II) in figure \ref{fig_discussion_I} and figure \ref{fig_discussion_II}, respectively.
In the figures, $\chi_I$ and $\chi_{II}$  at a certain cross-section of the filament are shown in the upper panels, and the resultant polarization profile is shown in the lower panels.

Figure \ref{fig_discussion_I} shows the map of $\chi_{I}$ on a $y-z$ plane, i.e., a cross-section of the filament. 
Here, we adopt the LOS of $(\theta, \phi)=(60^\circ, 0^\circ)$.
We choose the $x$ position of the cross-sections so that $\rfwhm$ is minimum (left) and maximum (right) (see the lower left panel of figure \ref{fig_scat_histo} for the distribution of $\rfwhm$).

The lower panels show the intensity ($I$) and polarized intensity ($PI$) profiles,
integrated along the vertical (LOS) direction of the cross sections shown in the upper panels.
The regions A, C, E, and G (regions framed in blue in the upper panels) show vertical speckling in green (positive $\chi_I$) and pink (negative $\chi_I$).
Vertical integration of such speckled regions results in depolarization.
It can be seen that the depolarization affects the profiles shown in the lower panels. Specifically, in the regions A, C, and G, $PI$ decreases compared to $I$.
In contrast, the regions B, D, and F (regions framed in red) are monochromatic green or pink, indicating no depolarization in the vertical (LOS) direction.
The figure shows that the depolarization due to the magnetic field fluctuations along the LOS can influence the polarized intensity.

Figure \ref{fig_discussion_II} shows the map and contours (solid) of $\chi_{I}$ on a $y-z$ plane, i.e., a cross-section of the filament. Here, we adopt the LOS of $(\theta, \phi)=(15^\circ, 0^\circ)$.
The $x$ positions of the cross-sections are chosen so that $\rfwhm$ is minimum (left) or maximum (right) along the filament (see the upper right panel of figure \ref{fig_scat_histo} for the distribution of $\rfwhm$).
The contours of the normalized density ($\rho/\rho_{\rm{max}}$) are also shown by dashed lines.
Both solid and dashed contours are normalized by their peak values.

If a solid contour is enclosed by a dashed contour of the same value (yellow-filled areas in the upper panels), the polarized emission becomes relatively weaker due to the alignment of $\boldsymbol{B}$ toward $\boldsymbol{l}$.
Conversely, when a solid contour encloses a dashed contour of the same value (red-filled areas in the upper panels), the polarized emission becomes relatively stronger.

The upper left panel of figure \ref{fig_discussion_II} shows that the contours of $\chi_{II}$ (solid) are horizontally more compact compared to the contours of normalized density (dashed),
leading to narrower $PI$ profiles than $I$. 
Meanwhile, the upper right panel reveals a dent in the solid contour near the peak of $I$ (around position E in the lower panel),
resulting in weaker polarized emission around the peak. 
Consequently, the $PI$ around the peak of $I$ (position E) decreases,
and the normalized $PI$ distribution appears to be broadened.
In this example, we conclude that process (II) plays a role in depolarization.

We examined $\chi_I$ and $\chi_{II}$ for almost the entire cross-section and various lines of sight, and unfortunately could not conclude that either process (I) or process (II) plays a dominant role. 
Rather, we found many instances where both process (I) and process (II) seem to affect the structures of the $PI$ profile.
This is probably because, when the global magnetic field projected onto the POS becomes small (when process (II) is at work), the turbulent eddies in the filament easily bend the magnetic field on the POS and more easily produce depolarization along the LOS (process (I)).

\section{Conclusion and discussion}
\label{sec_conclusion}
In this study, by combining 3D magnetohydrodynamics (MHD) simulations and synthetic observations, we investigate the properties of polarized emissions from turbulent filamentary molecular clouds.
In particular, we examine the scalar quantity $\rfwhm$, introduced by \citet{2021ApJ...923L...9D}, defined as the ratio of the Full Width at Half Maximum (FWHM) of the polarized intensity ($PI$) profile to that of the total intensity ($I$) along the minor axis of the filament.

Our results are summarized as follows:
\begin{itemize}
\item We find that, as the magnetic field inclination relative to the POS ($\theta$) increases, the mean ($\rbar$) and variance ($\rsigma$) of $\rfwhm$ within a filament decrease and increase, respectively. These trends are independent of the magnetic field strength, but they become less pronounced as the rotation angle $\phi$ increases.
\item If $\rbar$ is smaller than unity and/or $\rsigma$ is large in an observed filament, it suggests that the filament has a large inclination angle $\theta$ and a small rotation angle $\phi$ (see figure \ref{fig_def_angle} for definitions of the angles).
\item Both the depolarization due to the fluctuations of the magnetic field along the LOS by small-scale ($<0.1 \rm{pc}$) turbulence and the alignment of the global magnetic field toward the LOS by large-scale ($>0.1 \rm{pc}$) turbulence play roles in shaping the $PI$ profile.
\end{itemize}

Our research has certain limitations and simplifications.
Our models assume a spatially uniform dust grain alignment degree (constant $\epsilon_{\rm align}$).
However, according to Radiative Alignment Torques (RATs) theory \citep{2008MNRAS.388..117H} for example, the alignment degree is expected to decrease in dense inner regions of clouds due to the shielding of external radiation.
This may influence the $PI$ profile, especially in the inner regions of the cloud core.
We will investigate how the spatial variations of the dust alignment degree affect the polarization profiles and our conclusions in a future study.

The magnetic field strength of the filaments considered in this paper ($0.7-7\ \rm\mu G$) is relatively weak compared to the observations.
For example, \citet{2022MNRAS.510.6085L} shows the magnetic field strength in L1495/B211 is $10$-$80\ \rm\mu G$  depending on the regions, which is stronger than the value in our simulations.
We show that magnetic braking can be effective even with the moderate magnetic field strengths of $7.2\ \rm\mu G$ (see figure \ref{fig_jspe}).
Therefore, for more strongly magnetized filaments, the magnetic field is expected to play a more important role in the filament evolution.

However, simulating filaments with strong magnetic fields is challenging because the plasma $\beta$ becomes very small outside the filament, leading to extremely small timesteps and significantly increased computational costs.
Therefore, future studies are desired to explore well-designed density and magnetic field configurations.

In this paper, we have not discussed the required observational sensitivity to apply $\rfwhm$ analysis to observational studies. 
This is because we are particularly interested in the theoretical aspects of why the difference in the width of the $I$ and $PI$ profiles occurs and what it means.  
In actual observations, $PI$ is generally very weak compared to $I$.
Therefore, observational sensitivity is an important issue in applying $\rfwhm$ analysis in real observations.
If we look at the values of the color bars in figure \ref{fig_synob_overview} (a) and (b), we can see that the intensity of $PI$ is a few \% of that of $I$. 
Therefore, high sensitivity (e.g., $S/N\gtrsim 100$ ) would be required to obtain an accurate measurement of $\rfwhm$.
With current instruments such as JCMT and POL-2, it may be difficult to observe $\rfwhm$ except for limited filaments with favorable observation conditions (such as the filament observed in \citet{2021ApJ...923L...9D}).
However, future instruments (e.g., Atacama Large Aperture Submillimeter Telescope (atLAST)\footnote{$\langle$https://www.atlast.uio.no$\rangle$.} and Large Submillimeter Telescope (LST)\footnote{$\langle$https://www.lstobservatory.org$\rangle$}) will have higher sensitivities and we will be able to perform analyses presented in this study. In this sense, our finding of the variation of $\rfwhm$ and its relation to magnetic field inclination is important for future observations.

\begin{ack}
We thank Prof. Shu-ichiro Inutsuka, and Prof. Yasuo Doi for fruitful discussions and their encouragement. 
Numerical computations were performed on Cray XC50 at the Center for Computational Astrophysics, National Astronomical Observatory of Japan.
This study is supported by JST FOREST Program, Grant Number JPMJFR2234.
\end{ack}

\bibliographystyle{aasjournal}

\end{document}